\newcommand{\beq}{\begin{equation}}
\newcommand{\eeq}{\end{equation}}
\newcommand{\bea}{\begin{eqnarray}}
\newcommand{\eea}{\end{eqnarray}}
\newcommand{\non}{\nonumber}
\newcommand{\bfa}{{\bf (a)}}
\newcommand{\bfb}{{\bf (b)}}
\newcommand{\bfc}{{\bf (c)}}
\newcommand{\bfd}{{\bf (d)}}
\newcommand{\bfe}{{\bf (e)}}

\newcommand{\diag}{\mathrm{diag}}
\newcommand{\tr}{{\rm tr}}

\newcommand{\half}{{\textstyle {1\over 2}}}

\newcommand{\PP}{\mathrm{I}\kern -2.5pt \mathrm{P}}
\newcommand{\R}{\mathrm{I}\kern -2.5pt \mathrm{R}}
\newcommand{\sPP}{\mathrm{I}\kern -1.6pt \mathrm{P}}
\newcommand{\sR}{\mathrm{I}\kern -1.6pt \mathrm{R}}
\newcommand{\1}{1\kern -3pt \mathrm{l}}
\newcommand{\Z}{\mathsf{Z}\kern -4pt \mathsf{Z}}
\newcommand{\bet}{{b_0}}
\newcommand{\mm}{m}
\newcommand{\Ei}{e_i}
\newcommand{\Ri}{R_i}
\newcommand{\Rk}{R_k}
\newcommand{\Wnot}{W_0}
\newcommand{\Wpp}{\Wnot''(e_i)}
\newcommand{\Wzpp}{\Wnot''(\mm)}
\newcommand{\Wmat}{W_{\cN=2}}
\newcommand{\Lam}{\Lambda}
\newcommand{\del}{\partial}
\newcommand{\MxM}{M_0 \times M_0}
\newcommand{\tQ}{\tilde{Q}}
\newcommand{\tq}{\tilde{q}}
\newcommand{\om}{\omega}
\def\oms{ \omega_{\rm s} }
\def\omd{ \omega_{\rm d} }
\def\omrp{ \omega_{\rm rp} }
\def\free{ F_{\rm s} }
\def\fdisk{ F_{\rm d} }
\def\frp{ F_{\rm rp} }
\def\rp{ \R \PP^2} 
\newcommand{\D}{{\rm d}}
\newcommand{\TDz}{ \, T(z) \, \D z}
\newcommand{\omDz}{ \, \om(z) \, \D z}
\newcommand{\omsDz}{ \, \oms(z) \, \D z}
\newcommand{\cF}{\mathcal{F}}
\newcommand{\cG}{\mathcal{G}}
\newcommand{\cN}{\mathcal{N}}
\newcommand{\cO}{\mathcal{O}}
\newcommand{\cFpert}{\mathcal{F}_{\rm pert}}

\newcommand{\cFone}{\mathcal{F}_{1-\rm inst}}
\newcommand{\cFtwo}{\mathcal{F}_{2-\rm inst}}
\newcommand{\kk}{k}
\newcommand{\cFkinst}{\mathcal{F}_{\kk-\rm inst}}
\newcommand{\instsum}{2 \pi i \sum_{\kk=1}^\infty \kk \Lam^{\bet \kk} \cFkinst}
\newcommand{\insttwo}{2 \pi i \sum_{\kk=1}^\infty \kk \Lam^{2N \kk} \cFkinst}
\newcommand{\U}{\mathrm{U}}
\newcommand{\Sp}{\mathrm{Sp}}
\newcommand{\SO}{\mathrm{SO}}
\newcommand{\SU}{\mathrm{SU}}
\newcommand{\Acyc}{\oint_{A_i}}
\newcommand{\Bcyc}{\oint_{B_i}}
\newcommand{\Azcyc}{\oint_{A_0}}
\newcommand{\infcyc}{\oint_{\infty}}
\def\sumN{ \sum_{i=1}^{N} }

\def\ddSj{ \sum_{j=1}^N \frac{\del} {\del S_j}  }
\def\prodNf{ \prod_{I=1}^{N_f} }
\def\sumk{ \sum_{k\neq i} }
\def\sumell{ \sum_{\ell \neq i} }
\def\vev#1{ \langle {#1} \rangle }
\def\bvev#1{ \bigg\langle {#1} \bigg\rangle }

\def\tads{ \vev{\tr\, \Psi_{ii} }_{S^2} }
\def\tadd{ \vev{\tr\, \Psi_{ii} }_{\rm disk}   }
\def\tadrp{ \vev{\tr\, \Psi_{ii} }_{\sR \sPP^2}   }

\def\psiisqs{ \vev{\tr\, \Psi^2_{ii} }_{S^2} }
\def\psiisqd{ \vev{\tr\, \Psi^2_{ii} }_{\rm disk}   }
\def\psiisqrp{ \vev{\tr\, \Psi^2_{ii} }_{\sR \sPP^2}   }

\def\psizs{ \vev{\tr\, \Psi_{00} }_{S^2} }
\def\psizd{ \vev{\tr\, \Psi_{00} }_{\rm disk}   }
\def\psizrp{ \vev{\tr\, \Psi_{00} }_{\sR \sPP^2}   }

\def\psizsqs{ \vev{\tr\, \Psi^2_{00} }_{S^2} }
\def\psizsqd{ \vev{\tr\, \Psi^2_{00} }_{\rm disk}   }
\def\psizsqrp{ \vev{\tr\, \Psi^2_{00} }_{\sR \sPP^2}   }
\def\gs{ g_{s} }
\def\laSW{ \lambda_{SW} }
\def\Weff{ W_{\rm eff}}
\def\eij{  e_{ij} }
\def\eik{  e_{ik} }
\def\eil{  e_{i\ell} }
\def\ekl{  e_{k\ell} }
\def\aij{  a_{ij} }
\def\aik{  a_{ik} }
\def\akl{  a_{k\ell} }
\def\ail{  a_{i\ell} }
\def\vevS{ \bigg|_{\vev{S}}  }
\def\bigvevS{ \Bigg|_{\vev{S}}  }

\documentclass[12pt]{article}

\usepackage{graphicx}

\def\theequation{\thesection.\arabic{equation}}

\newdimen\tableauside\tableauside=1.0ex
\newdimen\tableaurule\tableaurule=0.4pt
\newdimen\tableaustep
\def\phantomhrule#1{\hbox{\vbox to0pt{\hrule height\tableaurule
width#1\vss}}}
\def\phantomvrule#1{\vbox{\hbox to0pt{\vrule width\tableaurule
height#1\hss}}}
\def\sqr{\vbox{%
  \phantomhrule\tableaustep
\hbox{\phantomvrule\tableaustep\kern\tableaustep\phantomvrule\tableaustep}%
  \hbox{\vbox{\phantomhrule\tableauside}\kern-\tableaurule}}}
\def\squares#1{\hbox{\count0=#1\noindent\loop\sqr
  \advance\count0 by-1 \ifnum\count0>0\repeat}}
\def\tableau#1{\vcenter{\offinterlineskip
  \tableaustep=\tableauside\advance\tableaustep by-\tableaurule
  \kern\normallineskip\hbox
    {\kern\normallineskip\vbox
      {\gettableau#1 0 }%
     \kern\normallineskip\kern\tableaurule}%
  \kern\normallineskip\kern\tableaurule}}
\def\gettableau#1 {\ifnum#1=0\let\next=\null\else
  \squares{#1}\let\next=\gettableau\fi\next}
\newcommand{\Yfund}{\tableau{1}}
\newcommand{\Ysymm}{\tableau{2}}
\newcommand{\Yasymm}{\tableau{1 1}}

\begin{document}

\begin{flushright} 
{\tt hep-th/0403129}\\ 
BRX-TH-536\\ 
BOW-PH-131 \\
\end{flushright}
\vspace{30mm} 
\begin{center}
{\bf\Large\sf 
Improved matrix-model calculation of 
the {\large $\cN=2$ } prepotential}

\vskip 5mm
Marta~G\'omez-Reino\footnote{Research 
supported by the DOE under grant DE--FG02--92ER40706.}$^{,a}$,
Stephen G. Naculich\footnote{Research
supported in part by the NSF under grant PHY-0140281.}$^{,b}$,
and Howard J. Schnitzer\footnote{Research
supported in part by the DOE under grant DE--FG02--92ER40706.\\
{\tt \phantom{aaa} marta,schnitzer@brandeis.edu; naculich@bowdoin.edu}\\
}$^{,a}$

\end{center}

\vskip 1mm

\begin{center}
$^{a}${\em Martin Fisher School of Physics\\
Brandeis University, Waltham, MA 02454}

\vspace{.2in}

$^{b}${\em Department of Physics\\
Bowdoin College, Brunswick, ME 04011}
\end{center}

\vskip 2mm

\begin{abstract} 

We present a matrix-model expression for the sum of 
instanton contributions to the prepotential of
an $\cN=2$ supersymmetric $\U(N)$ gauge theory,
with matter in various representations.
This expression is derived by combining the 
renormalization-group approach to the gauge theory 
prepotential with matrix-model methods.
This result can be evaluated
order-by-order in matrix-model perturbation theory
to obtain the instanton corrections to 
the prepotential.
We also show, using this expression, that the one-instanton
prepotential assumes a universal form.

\end{abstract}

\vfil\break

\setcounter{equation}{0}
\section{Introduction}
\label{sec:intro}

Over the past two years, many rich connections between 
supersymmetric gauge theories and matrix models have
been uncovered, beginning with the work
of Dijkgraaf and Vafa \cite{Dijkgraaf:2002,Dijkgraaf:2002dh}.
(For a review and list of references, see Ref.~\cite{Argurio:2003ym}.)
One aspect of this is the relation between matrix models
and the Seiberg-Witten solution of $\cN=2$ gauge 
theories \cite{Seiberg:1994},
which was elucidated for gauge group $\U(N)$ 
with matter in various representations
in Refs.~\cite{Dijkgraaf:2002dh,Dijkgraaf:2002pp,
Naculich:2002hi,Klemm:2002pa,Naculich:2002hr,Naculich:2003ka,Boels:2003}
and for other gauge groups in 
Ref.~\cite{AhnNam:2003}.

In Refs.~\cite{Naculich:2002hi,Naculich:2002hr,Naculich:2003ka},
the matrix-model approach was used to compute
the one-instanton contribution to the $\cN=2$ prepotential
for $\U(N)$ gauge theories with matter in fundamental,
symmetric, or antisymmetric representations,
with the results in agreement with previous calculations 
made within the Seiberg-Witten 
framework. 
The matrix-model calculation requires 
the evaluation of the free energy to two-loops in perturbation theory
(together with a tadpole calculation in the matrix model
to relate the classical moduli $e_i$ to the 
quantum moduli $a_i$ of Seiberg-Witten theory).
The final expressions for the one-instanton prepotential,
however, are much simpler than the intermediate calculations.
Moreover, it was observed {\it a posteriori} 
(in sec.~6 of Ref.~\cite{Naculich:2003ka})
that the one-instanton prepotential 
for all the theories considered takes the universal form
\beq
\label{eq:introuniv}
2 \pi i \Lam^\bet \cFone
=  \sum_i  \frac{\vev{S_i}}{\Wpp} + \frac{\vev{S_0}}{\Wzpp} 
\eeq
where the second term is present only in theories containing an
antisymmetric hypermultiplet (with mass $\mm$).
Here $\Wnot'(z) = \alpha \prod_{i=1}^N (z-e_i)$,
$\vev{S_i}$ and $\vev{S_0}$ are the values of the
glueball fields $S_i$ and $S_0$ 
that extremize the effective superpotential,
and $\Lam^\bet$ is the one-instanton scale.
Particularly intriguing is that 
the r.h.s.~of eq.~(\ref{eq:introuniv}) 
can be evaluated using a one-loop
calculation in the matrix model,
despite the fact that two-loop calculations were 
required to arrive at this result.
This strongly suggests that a much simpler route to the prepotential exists.

In this paper, we provide this simpler approach
to the prepotential by combining matrix-model methods 
with the renormalization-group approach
to the $\cN=2$ prepotential \cite{Matone:1995,
D'Hoker:1997ph,D'Hoker:2002pn,Gomez-Reino:2003cp}.
We obtain a general expression 
for the sum of the instanton contributions to the prepotential 
in terms of matrix-model quantities:
\bea
\label{eq:introthreeterms}
\instsum&=& \half  \sumN  
\Bigg[ \ddSj \gs \psiisqs     +  \psiisqd  + 4\, \psiisqrp \Bigg]\bigvevS 
\non \\
 &+&  
\half \Bigg[ \ddSj \gs \psizsqs  + \psizsqd  + 4\, \psizsqrp \Bigg] \bigvevS
\\
&-& \half \sumN  \Bigg[ 
 \ddSj \gs \tads  + \tadd  + 4\, \tadrp \Bigg]^2\bigvevS \, .\non
\eea
The r.h.s.~can be evaluated order-by-order 
in matrix-model perturbation theory.
The one-instanton prepotential
requires only a one-loop matrix-model calculation,
rather than the two-loop calculation used previously,
and yields the universal expression (\ref{eq:introuniv}).
It is also easier to use eq.~(\ref{eq:introthreeterms})
to obtain higher-instanton contributions to the prepotential,
as we illustrate for the case of the pure $\cN=2$ $\U(N)$ theory.

Section 2 contains a review of the salient features
of the Seiberg-Witten and matrix-model approaches 
to $\cN=2$  supersymmetric gauge theories.
Section 3 contains the main result of our paper, 
the derivation of eq.~(\ref{eq:introthreeterms}).
In Sec.~4, we evaluate this expression for all $\U(N)$ gauge theories
with asymptotically-free matter content to obtain the 
one-instanton correction to the prepotential. 
We also show that eq.~(\ref{eq:introuniv}) is valid
for pure $\cN=2$ $\SO(N)$ and $\Sp(N)$ theories.
In Sec.~5, we compute the two-instanton prepotential
for pure $\cN=2$ $\U(N)$  gauge theory,
using a two-loop matrix-model calculation,
one loop fewer than would be required by the method of 
Ref.~\cite{Naculich:2002hi}.
Section 6 contains our conclusions.

\setcounter{equation}{0}
\section{$\cN=2$ supersymmetric gauge theory}
\label{sec:gauge}

In this section, we review $\cN=2$ supersymmetric gauge theory,
first in the Seiberg-Witten approach, and then
in the matrix-model approach.
Specifically, we will treat
all classes of asymptotically-free $\cN=2$ $\SU(N)$ gauge theories
with or without matter:
\\[5pt]
\noindent \bfa\ 
$\cN=2$ $\SU(N)$ with  
$N_f$ $\Yfund$ hypermultiplets 
($0 \le  N_f < 2N$),
\vphantom{$\Yasymm$}
\\[3pt] \noindent 
\bfb\   $\cN=2$ $\SU(N)$ with one
$\Ysymm$ and $N_f$ $\Yfund$ hypermultiplets 
($0 \le N_f < N-2$),
 \vphantom{$\Yasymm$}
\\[3pt] \noindent 
\bfc\  $\cN=2$ $\SU(N)$ with one
$\Yasymm$ and 
$N_f$ $\Yfund$ hypermultiplets ($0 \le N_f < N+2$),
\\[3pt] \noindent 
\bfd\ 
$\cN=2$ $\SU(N)$ with two
$\Yasymm$ and 
$N_f$ $\Yfund$ hypermultiplets 
($0 \le N_f < 4$).  

\subsection{Seiberg-Witten approach}

In the Seiberg-Witten (SW) approach \cite{Seiberg:1994},
the Coulomb branch of an $\cN=2$ gauge theory
is described in terms of an algebraic curve $\Sigma$
and meromorphic differential $\laSW$,
whose explicit forms depend on the gauge group and matter 
content \cite{Seiberg:1994,
Klemm:1995,Landsteiner:1998ei,Landsteiner:1998pb}.
Consider a generic point of the Coulomb branch, 
with the $\SU(N)$ gauge symmetry broken down to $\U(1)^{N-1}$.
The SW differential can be expressed
as $\laSW = z \TDz$,
where\footnote{\label{foot:norm}Throughout this paper, 
a factor of $1/2 \pi i$ is implied in any expression involving $\oint$.}
\beq
\label{eq:Tdef}
\Acyc \TDz = 1, \qquad\qquad \Bcyc \TDz = 0
\eeq
with  $\{ A_i, B_i \}$ a canonical basis of homology cycles on $\Sigma$.
The quantum moduli $a_i$ and their duals $a_{D,i}$ 
are defined as periods of the SW differential
\beq
\label{eq:adef}
a_i = \Acyc z \TDz,   \qquad\qquad  a_{D,i} = \Bcyc  z \TDz \, .
\eeq
The gauge theory prepotential $\cF(a)$  
and 
the matrix $\tau_{ij}$ of gauge couplings of the unbroken gauge group
are given by
\beq
\label{eq:taudef}
a_{D,i} = {\partial  \cF(a) \over {\partial a_i}},
\qquad\qquad
\tau_{ij} =        {\partial  a_{D,i} \over  \partial a_j}
          =        {\partial^2 \cF(a) \over  \partial a_i \partial a_j}\, .
\eeq
The SW curves and differentials for theories \bfa--\bfd,
whose explicit forms will not be needed in this paper, 
depend on a set of classical moduli $\Ei$,
the hypermultiplet masses,
and the quantum scale $\Lam$ of the gauge theory.
By expanding $\laSW$ in powers of $\Lam$,
one may compute the periods $a_i$ and $a_{D,i}$ 
in a weak-coupling expansion. 
To lowest order, $a_i$ coincides with the classical moduli:
$a_i = \Ei + \cO(\Lam^\bet)$,
where $ \bet =  2N - \sum_n I(R_n)$
with $I(R_n)$ the Dynkin indices of the representations $R_n$
of the matter hypermultiplets\footnote{For the theories considered 
in this paper, 
$\bet=2N-N_f$, $N-2-N_f$, $N+2-N_f$, and $4-N_f$ for 
cases ${\bf (a),\,(b),\,(c),}$ and ${\bf (d)}$ respectively.}.
Next, $a_{D,i}$ may be integrated with respect to $a_i$ 
to obtain the prepotential in the form
\beq
\label{eq:prepot}
\cF(a)  
= \cFpert(a, \log \Lam)  
+ \sum_{\kk=1}^\infty \Lam^{\bet \kk} \cFkinst  (a)
\eeq
where $ \cFpert(a, \log \Lam) $
consists of the classical and one-loop prepotential,
and $\cFkinst (a) $ is the $\kk$-instanton contribution.
Using the Seiberg--Witten approach, the instanton 
corrections to the prepotential for theory \bfa\ were computed in 
Refs.~\cite{D'Hoker:1997ph,D'Hoker:1997nv,Chan:1999gj,whitham}. 
(More recent approaches to computing the prepotential
may be found in Refs.~\cite{Nekrasov:2002qd,Flume:2002}.) 
For theories \bfb\ and \bfc, the one-instanton prepotentials
were computed in Refs.~\cite{Naculich:1998,Ennes:1998ve}.
For theory \bfd, 
the conjectured form of $\cFone$ was used in Ref.~\cite{Ennes:1999rn}
to reverse-engineer the (approximate) form of the SW curve 
for the theory; 
proposals for the exact curve 
were made in Ref.~\cite{Ksir:2002}, 
but not in a form that enabled the extraction of $\cFkinst$.

\subsection{Matrix model approach}
\label{sec:MMapproach}

We now sketch the 
relation between $\cN=2$ supersymmetric gauge theories 
and large $M$ matrix models,
referring to 
Refs.~\cite{Dijkgraaf:2002pp,Naculich:2002hi,Naculich:2002hr,Naculich:2003ka}
for further details.

Consider an $\cN=2$ $\U(N)$ gauge theory\footnote{ The $\SU(N)$ gauge 
group appears naturally within the Seiberg--Witten context, whereas the 
gauge group $\U(N)$ appears naturally in the matrix model context. 
The $\SU(N)$ gauge theory discussed in the previous subsection is the
non-trivial piece of this $\U(N)$ theory.}
containing an $\cN=2$ adjoint vector multiplet 
(which includes an $\cN=1$ chiral superfield $\phi$)
as well as $\cN=2$ matter hypermultiplets in various representations
(each of which comprises a pair of $\cN=1$ chiral superfields $q$ and $\tq$).
The superpotential for the gauge theory is taken to be
$\Wnot (\phi) + \Wmat (\phi,q,\tq) $,
where $ \Wmat (\phi, q, \tq)$ is the $\cN=2$ superpotential
(including possible mass terms for the matter hypermultiplets) 
and $\Wnot'(z) = \alpha \prod_{i=1}^N (z-e_i)$.
The inclusion of $\Wnot(\phi)$ in  the superpotential
breaks\footnote{Upon the conclusion of 
the matrix-model computation, we take $\alpha \to 0$, 
restoring $\cN=2$ supersymmetry.}
the supersymmetry to $\cN=1$,
and freezes the moduli of the Coulomb branch of vacua to
$\vev{\phi} = \diag (e_1, e_2, \ldots, e_N)$,
breaking the $\U(N)$ gauge symmetry to $\U(1)^N$.

More general vacua (with $\vev{q}$,$\vev{\tq}  \neq 0$) are possible
when matter hypermultiplets are present.
In particular, for theories
with symmetric and antisymmetric representations,
there are vacua which leave $\Sp(N_0)$ or $\SO(N_0)$ 
factors in the unbroken gauge group \cite{Klemm:2003cy}.   
Thus, for theory \bfb, containing a symmetric hypermultiplet, 
the generic vacuum state has unbroken gauge group 
$\SO(N_0) \times \prod \U(N_i)$;
for theory \bfc, containing an antisymmetric hypermultiplet,
the generic vacuum state has unbroken gauge group 
$\Sp(N_0) \times \prod \U(N_i)$;
for theory \bfd, containing two antisymmetric hypermultiplets,
the generic vacuum state has unbroken gauge group 
$\Sp(N_0) \times \Sp(N'_0) \times \prod \U(N_i)$.
On the Coulomb branch, one has $N_i = 1$ for $i=1,\cdots, N$
and $N_0=N'_0=0$.
It turns out that the apparently trivial $\Sp(0)$ factors
(but not the $\SO(0)$ factors)
play an important 
role \cite{Aganagic:2003xq,Cachazo:2003kx,AhnFeng:2003,Intriligator:2003xs}
in the IR dynamics 
and in the matrix-model correspondence,
as we will see below.

The $\U(N)$ gauge theory just described 
(with supersymmetry broken to $\cN=1$)
is related to a $\U(M)$ matrix 
model \cite{Dijkgraaf:2002, Dijkgraaf:2002dh}
in which each $\cN=1$ chiral superfield of the
gauge theory has a matrix-model analog:
specifically, an $M \times M$ matrix $\Phi$ (corresponding to $\phi$)
together with $Q$ and $\tQ$, 
which are vectors or symmetric/antisymmetric matrices 
depending on the representation of the
corresponding matter superfields $q$ and  $\tq$.
The superpotential of the gauge theory is taken to be 
the potential of the matrix model, whose partition function is
\beq
\label{eq:partfcn}
Z = 
 \frac{1}{\mathrm{vol}(G)}
\int \D\Phi \, \D Q \, \D \tQ\,
\exp \left( 
-{1 \over \gs} \left[  \Wnot(\Phi) + \Wmat (\Phi,Q,\tQ)  \right]
\right)
\eeq
where $G$ is the unbroken gauge group of the matrix model.

The partition function (\ref{eq:partfcn})
is evaluated perturbatively about the extremum
\beq
\label{eq:Phinought}
\Phi_0 = 
\pmatrix{ 
\mm \1_{M_0}& 0& \cdots& 0 \cr
0& e_1 \1_{M_1}& \cdots& 0 \cr
\vdots& \vdots& \ddots& \vdots \cr
0& 0& \cdots&  e_N \1_{M_N}  
},\qquad\qquad M= M_0 + \sum_{i=1}^N M_i
\eeq
where the $\MxM$ block (with $M_0$ even) is included only when 
the gauge theory contains an antisymmetric hypermultiplet
(of mass $\mm$).   
In general, $Q$ and $ \tQ$ vanish at the extremum,
except when the theory contains an antisymmetric hypermultiplet,
in which case $Q_0$ and $\tQ_0$ contain $\MxM$ blocks
proportional to the symplectic unit $J$.
When the $\MxM$ block is absent, 
the vacuum state (\ref{eq:Phinought}) breaks the $\U(M)$ symmetry 
of the matrix model to $G = \prod_{i=1}^N  \U(M_i)$.
When the $\MxM$ block is present,
the vacuum state (\ref{eq:Phinought}) 
(together with $Q_0$ and $\tQ_0$) breaks the $\U(M)$ symmetry 
of the matrix model to $G = \Sp(M_0) \times \prod_{i=1}^N  \U(M_i)$.

The presence of the additional $\MxM$ block for theories
containing an antisymmetric representation \bfc,
and its absence for theories containing a symmetric representation \bfb,
was discovered to be necessary to obtain the correct SW curve,
differential, and one-instanton prepotential from the matrix model
in Ref.~\cite{Naculich:2003ka}.
This is an example of a more general phenomenon
in which the correct IR description of the gauge theory 
requires \cite{Aganagic:2003xq,Cachazo:2003kx,AhnFeng:2003,Intriligator:2003xs}
the inclusion of a glueball field 
(in this case $S_0 = \gs M_0$) corresponding to 
apparently trivial $\Sp(0)$ factors in the unbroken gauge group.
Such a prescription resolves \cite{Cachazo:2003kx,Intriligator:2003xs}
an apparent discrepancy
\cite{Kraus:2003jf,Alday:2003gb,Kraus:2003jv}
in a related $\cN=1$ $\Sp(N)$ gauge theory.

For theories with two antisymmetric hypermultiplets \bfd,
the unbroken gauge theory symmetry on the Coulomb 
branch is $\Sp(0) \times \Sp(0) \times \U(1)^N $. 
The prescription in Ref.~\cite{Intriligator:2003xs}
requires the inclusion of two glueball fields $S_0$ and $S'_0$
for the two $\Sp(0)$ factors.
Correspondingly, the perturbative matrix model 
must be expanded about an extremum 
that includes two additional blocks,
$\mm \1_{M_0}$ and $\mm' \1_{M'_0}$,
in $\Phi_0$ 
(and corresponding blocks, proportional to $J$, in $Q_0$ and $\tQ_0$),
breaking the matrix-model symmetry to 
$G = \Sp(M_0) \times \Sp(M'_0) \times \prod_{i=1}^N  \U(M_i)$.

To evaluate the partition function 
(\ref{eq:partfcn}) perturbatively about eq.~(\ref{eq:Phinought})
we write $\Phi= \Phi_0 + \Psi$.
A gauge choice \cite{Dijkgraaf:2002pp,Naculich:2002hi} allows us to take 
$\Psi$ to be block-diagonal,
$\Psi = \diag (\Psi_{00}, \Psi_{11},  \cdots, \Psi_{NN})$.
The matrix-model expectation values
$\vev{ \tr\, \Phi^n}$ can also be evaluated perturbatively
\beq
\label{eq:sumblocks}
\vev{ \tr\, \Phi^n}
=\sum_{i=1}^N \vev{ \tr_{\rm i}\, \Phi^n}  +\vev{ \tr_{\rm 0}\, \Phi^n} 
= \sum_{i=1}^N \vev{ \tr (e_i \1_{M_i} + \Psi_{ii})^n}
             + \vev{ \tr (\mm \1_{M_0} + \Psi_{00})^n}
\eeq
by computing matrix-model Feynman diagrams
containing insertions of $\Psi^m_{ii}$
(and $\Psi^m_{00}$, for theories with an antisymmetric representation).

Alternatively, one may evaluate the partition function
(\ref{eq:partfcn}) using saddle-point methods
(for a review, see Ref.~\cite{DiFrancesco:1995nw})
in which case one starts from the resolvent 
of the matrix model
\beq
\label{eq:resolv}
\om(z) \equiv  \gs \bvev{ \!\tr\!\left(\frac{1}{z-\Phi}\right) \!}
= \sum_{n=0}^{\infty} z^{-n-1} \gs \vev{ \tr\, \Phi^n},
\qquad\qquad
\gs \vev{ \tr\, \Phi^n} = \infcyc  z^n \omDz
\eeq
where $\infcyc$ represents a large contour in the $z$-plane,
taken counterclockwise.\footnote{See footnote \ref{foot:norm}.}
The resolvent $\om(z)$ classically has poles at $e_i$ 
(and $\mm$, if an antisymmetric hypermultiplet is present). 
In the saddle-point approximation, these open up into cuts,
and $\om (z)$ becomes a function on a multisheeted Riemann surface.
Let $A_i$ (and $A_0$) be cycles surrounding these cuts
on the first sheet.
The contour in eq.~(\ref{eq:resolv}) may be deformed into
a sum of cycles $A_i$, $A_0$,
and we may identify the integral around the cut at $e_i$  (or $\mm$)
with the trace over the $i$th (or $0$th) block of $\Phi$:
\beq
\gs \vev{\tr_{\rm i} \Phi^n} = \Acyc  z^n \omDz , \qquad\qquad
\gs \vev{\tr_{0} \Phi^n}     = \Azcyc  z^n \omDz 
\eeq
of which the $n=0$ case is 
\beq
\label{eq:Sdef}
S_i \equiv \gs M_i = \Acyc \omDz, \qquad\qquad
S_0 \equiv \gs M_0 = \Azcyc \omDz \, .
\eeq
It follows straightforwardly from eq.~(\ref{eq:sumblocks}) that
\beq
\label{eq:insertions}
\gs \vev{\tr\, \Psi^n_{ii} } = \Acyc  (z-e_i)^n \omDz , 
\qquad\qquad 
\gs \vev{\tr\, \Psi^n_{00} } = \Azcyc  (z-\mm)^n \omDz
\eeq
relating perturbative diagrams
(with insertions of $\Psi^n_{ii}$ or $\Psi^n_{00}$)
in the matrix model
to certain moments of the resolvent.

The next step is to express matrix-model quantities in a 
large $M$ (small $\gs$) expansion, with $S_i$, $S_0$ fixed.
The free energy can be written as
\beq
\label{eq:topol}
\log Z =  \sum_{\chi \le 2} \gs^{-\chi} F_{\chi} (S).
\eeq
Using 't Hooft double-line notation for the connected
diagrams of the matrix model,
this corresponds to the usual topological expansion \cite{'tHooft:1974jz}
characterized by $\chi$, 
the Euler characteristic $2 {-} 2g {-} h {-} q$
of the surface in which the Feynman diagrams are embedded
(where $g$ is the number of handles, 
$h$ the number of boundary components, 
and $q$ the number of crosscaps).
In the large $M$ limit,
the dominant contribution 
$\free \equiv \gs^2 \log  Z \big|_{\rm sphere} $
arises from planar diagrams that can be drawn on a sphere ($\chi=2$).
Theories with fundamental representations
give rise to surfaces with boundaries;
the dominant boundary contribution 
$\fdisk \equiv \gs \log  Z \big|_{\rm disk} $
comes from planar diagrams on a disk ($\chi=1$).
Theories with symmetric or antisymmetric representations
contain nonorientable surfaces; 
the dominant nonorientable contribution
$\frp \equiv \gs \log  Z \big|_{\sR \sPP^2} $ 
comes from planar diagrams on 
$\rp$, a sphere with one crosscap ($\chi=1$).
Hence,
\beq
\log Z =  \gs^{-2} \free + \gs^{-1} \left( \fdisk + \frp \right) + \cO(\gs^0)
\, .
\eeq
The resolvent can also be written in a large $M$ expansion:
\beq
\om = \oms + \gs (\omd +  \omrp ) + \cO(\gs^2)
\, .
\eeq
The perturbative diagrams contributing to 
the vevs $\vev{\tr\, \Psi^n_{ii} }$ and $\vev{\tr\, \Psi^n_{00} }$
may be classified by the surface on which they can be drawn,
thus
\beq
\label{eq:sphinsertions}
\gs \vev{\tr\, \Psi^n_{ii} }_{S^2} = \Acyc  (z-e_i)^n \omsDz ,   
\qquad
\gs \vev{\tr\, \Psi^n_{00} }_{S^2} = \Azcyc  (z-\mm)^n \omsDz, 
\qquad n \ge 1
\eeq
and similarly for the disk and $\rp$ contributions.

One last ingredient,
the effective superpotential,
is necessary to make the connection between
the matrix model and the gauge theory.
When the $\U(N)$ symmetry of the gauge theory 
is broken to $\prod_{i=1}^N \U(N_i)$, with $N_i=1$,
the superpotential takes the 
form \cite{Dijkgraaf:2002dh,Argurio:2002}
\beq
\label{eq:super}
\Weff=\, - \left[ \ddSj  \free  + \fdisk  + 4\, \frp\right] \, .
\eeq
In the matrix model, the parameters $S_i$ (and  $S_0$) are arbitrary, 
but extremizing (\ref{eq:super})
yields specific values $\vev{S_i}$ (and $\vev{S_0}$).
Specializing the Riemann surface 
on which the resolvent (\ref{eq:resolv}) is defined 
to $\vev{S}$ 
then yields the SW curve $\Sigma$ of the gauge theory.
The SW differential $\laSW = z \TDz$ is given by the matrix-model 
expression \cite{Naculich:2002hr,Kraus:2003jv,Gopakumar:2002wx,Naculich:2003cz}
\beq
\label{eq:MMTdef}
T(z) = \Bigg[ \ddSj  \oms  + \omd + 4\, \omrp \Bigg] \bigvevS
\eeq
also evaluated at the extrema $\vev{S}$ of the effective superpotential.
Finally, for $\U(N)$ with (or without) fundamental hypermultiplets,
the matrix of unbroken gauge couplings (\ref{eq:taudef}) is given 
by \cite{Dijkgraaf:2002dh}
\beq
\tau_{ij} = {1\over 2\pi i} \, 
{ \partial^2 \free \over \partial S_i \partial S_j} \vevS
\eeq
evaluated at $\vev{S}$. 
The prescription for $\tau_{ij}$ is modified 
for theories with symmetric
and antisymmetric representations
\cite{Naculich:2003ka,Naculich:2003cz}.
The $\cN=2$ prepotential (\ref{eq:taudef})
is then obtained \cite{Naculich:2002hi,Naculich:2002hr,Naculich:2003ka}
by integrating $\tau_{ij}$ twice with respect to $a_i$.

An equivalent approach \cite{Cachazo:2003kx,Cachazo:2003yc,Casero:2003gr}
is to define $\vev{S}$,
not as the extrema of an effective superpotential,
but as those values which ensure that $\TDz$,
defined by eq.~(\ref{eq:MMTdef}),
obeys eq.~(\ref{eq:Tdef}).
In cases in which the unbroken gauge group contains 
``trivial'' $\Sp(0)$ factors,
such as the theories considered in this paper with antisymmetric
hypermultiplets, one must further impose
$\Azcyc \TDz  = N_0 = 0 $.
Cachazo \cite{Cachazo:2003kx} emphasizes 
(in a related theory)
that the vanishing of the $A_0$ period of $\TDz$ does not
imply the vanishing of $S_0 = \Azcyc \omDz$.
In turn, the nonvanishing of $\Azcyc \omDz$
implies the presence of an $\MxM$ block in $\Phi_0$.

\section{Matrix-model computation of the ${\cal N}=2$ prepotential}
\label{sec:newstuff}

In this section, we derive a new matrix-model expression 
for the instanton contributions to the prepotential.
This simplifies the matrix-model computation of the one- and two-instanton
prepotentials, 
and allows us to prove that $\cFone$ has the simple
universal form (\ref{eq:introuniv}) 
presented in the introduction.

An efficient method for computing $\cFkinst$
for the $\SU(N)$ gauge theory 
with or without fundamental representations 
was developed within the Seiberg-Witten approach 
in Refs.~\cite{D'Hoker:1997ph,Chan:1999gj}
by making use of a renormalization-group equation
satisfied by the $\cN=2$ 
prepotential \cite{Matone:1995,D'Hoker:1997ph}. 
This method  was extended to $\SU(N)$ gauge theories 
with one symmetric or one antisymmetric representation 
in Refs.~\cite{D'Hoker:2002pn,Gomez-Reino:2003cp},
and can easily be generalized to include all cases in \bfb\ and \bfc\ above.
In all these cases, the renormalization-group equation 
for the prepotential takes the form 
\beq
\label{eq:RGfull}
2 \pi i \Lam \frac{\del \cF}{\del \Lam} 
= \bet u_2 + \cdots
\eeq
where $\cdots$ represents moduli-independent constants,
which would not affect the computation of $a_{D,i}$ or $\tau_{ij}$.
Here $u_2$ is defined as
\beq
\label{eq:udef}
u_2 = \oint_{\infty}  \half z^2 \TDz \,  .
\eeq
The perturbative (one-loop) prepotential satisfies 
\beq
\label{eq:RGpert}
2 \pi i \Lam \frac{\del \cFpert}{\del \Lam} 
= {\bet \over 2 }  \sumN a_i^2 + \cdots
\eeq
where $ \cdots $ represents $a_k$-independent constants.
(In writing this, we have assumed $\sum_i a_i = 0$,
as is appropriate for an $\SU(N)$ gauge theory.) 
Subtracting eq.~(\ref{eq:RGpert}) from (\ref{eq:RGfull}), 
and neglecting henceforth the moduli-independent constants,
we obtain
\beq
\label{eq:instsum}
\instsum = 
u_2 + \sumN \Ei (\half \Ei - a_i) - \half \sumN (a_i - \Ei)^2  
\eeq
where we have introduced the classical moduli $\Ei$.
Different choices for $\Ei$ are possible,
differing at $\cO(\Lam^\bet)$,
but the prepotential $\cF(a)$,
computed using eq.~(\ref{eq:taudef}),
is independent of this choice \cite{D'Hoker:1997nv}.
Equation (\ref{eq:instsum}) is obviously valid for any choice of $\Ei$,
so we will make a convenient choice, 
letting  $\Ei$ be the parameters that appear 
naturally in the matrix-model approach (sec.~\ref{sec:MMapproach}).

We now use eqs.~(\ref{eq:Tdef}) and (\ref{eq:adef}) 
to rewrite each of the terms in eq.~(\ref{eq:instsum})
as a contour integral, 
which will be suitable for re-expressing the equation
in terms of matrix-model quantities.
The sheet on which eq.~(\ref{eq:udef})
is defined has cuts centered (approximately) on $z=e_i$.
For theories containing an antisymmetric representation (of mass $m$),
there is also a cut\footnote{For theories \bfb\ and \bfc,
we are using the transformed cubic curves
(given, when $N_f=0$, 
by eqs.~(2.7) of Ref.~\cite{Naculich:2003ka}
or eq.~(7.7) of Ref.~\cite{Landsteiner:1998ei})
rather than the original cubic curves
(eqs.~(2.1) of Ref.~\cite{Naculich:2003ka})
derived from M-theory in Ref.~\cite{Landsteiner:1998ei}.
The transformed curves are the ones that emerge 
naturally in the matrix-model 
approach \cite{Naculich:2003ka,Klemm:2003cy,Naculich:2003cz}.
For the form of the transformed curves
for one $\Yasymm$ (or $\Ysymm$)  and $N_f>0$ $\Yfund$
hypermultiplets, see the appendix.}
centered (approximately) on $z=m$. 
The contour in eq.~(\ref{eq:udef}) may be deformed to give
\beq
\label{eq:utwo}
u_2 = \sumN \Acyc \half z^2 \TDz
    + \Azcyc \half z^2 \TDz\,.
\eeq
The first term in eq.~(\ref{eq:utwo})
combines with the second term on the r.h.s.~of eq.~(\ref{eq:instsum})
to give
\beq
\sumN \Acyc \half z^2 \TDz
+ \sumN \Acyc \Ei (\half \Ei - z) \TDz
= \sumN \Acyc \half (z- \Ei)^2 \TDz\,.
\eeq
Collecting all the contributions, we obtain
\bea
\label{eq:threeterms}
\instsum &=& \cG_1 + \cG_2 + \cG_3\, ,  \\
\cG_1 &=& \sumN \Acyc \half (z- \Ei)^2 \TDz \, ,\non\\
\cG_2 &=&  \Azcyc \half  z^2  \TDz \, ,  \non\\
\cG_3 &=& - \half \sumN  \left[ \Acyc (z - \Ei) \TDz \right]^2  \non
\eea
where $\cG_2$ is present only in theories containing
an antisymmetric hypermultiplet,
and $\cG_3$ only contributes to two-instantons 
and higher.

Even purely within the context of SW theory, 
eq.~(\ref{eq:threeterms}) is a useful equation for
computing the instanton contributions to the prepotential
by expanding $\TDz$ in powers of $\Lam$.
However,  our interest is in recasting this expression
in terms of matrix-model quantities.
Using eqs.~(\ref{eq:MMTdef}) and (\ref{eq:sphinsertions}) etc.,
we obtain
\bea
\label{eq:threetermsMM}
\cG_1 &=& \half  \sumN  
\Bigg[ \ddSj \gs \psiisqs     +  \psiisqd  + 4\, \psiisqrp \Bigg]\bigvevS 
\, ,\non \\
\cG_2 &=&  
\half \Bigg[ \ddSj \gs \psizsqs  + \psizsqd  + 4\, \psizsqrp \Bigg] \bigvevS
\, , \\
\cG_3 &=& - \half \sumN  \Bigg[ 
 \ddSj \gs \tads  + \tadd  + 4\, \tadrp \Bigg]^2\bigvevS  \non
\eea
which is the main result of this paper.
In writing this, we have used 
\bea
\Azcyc \half  z^2  \TDz 
&=& 
\Azcyc \half  m^2 \TDz  \\
&+&
m \Bigg[ \ddSj \gs \psizs  + \psizd  + 4\, \psizrp \Bigg] \bigvevS\non\\
&+& 
\half \Bigg[ \ddSj \gs \psizsqs  + \psizsqd  + 4\, \psizsqrp \Bigg] \bigvevS
\non
\eea
where  the first two terms on the r.h.s.~vanish because
$\Azcyc \TDz = 0$ and 
$\psizs = \psizd  = \psizrp =0$ since $\Psi_{00} \in \Sp(M_0)$  
\cite{Naculich:2003ka}.

In the following two sections, we will evaluate the 
expression (\ref{eq:threetermsMM}) for asymptotically-free
$\cN=2$ $\U(N)$ gauge theories.

\section{Universality of the one-instanton prepotential}

In this section, we will demonstrate that the one-instanton 
contribution to the prepotential
has a simple universal form for all the theories considered
in this paper (and for others as well).
The only terms in eq.~(\ref{eq:threetermsMM})
that contribute to the one-instanton prepotential are
\beq 
2 \pi i \Lam^\bet \cFone
=   \half \bigg[ \sumN  \ddSj \gs \psiisqs   + 4\, \psizsqrp \bigg] \vevS \, .
\eeq

\begin{figure}[htbp]
\begin{center}
\includegraphics[width=8.0cm,height=3.0cm]
{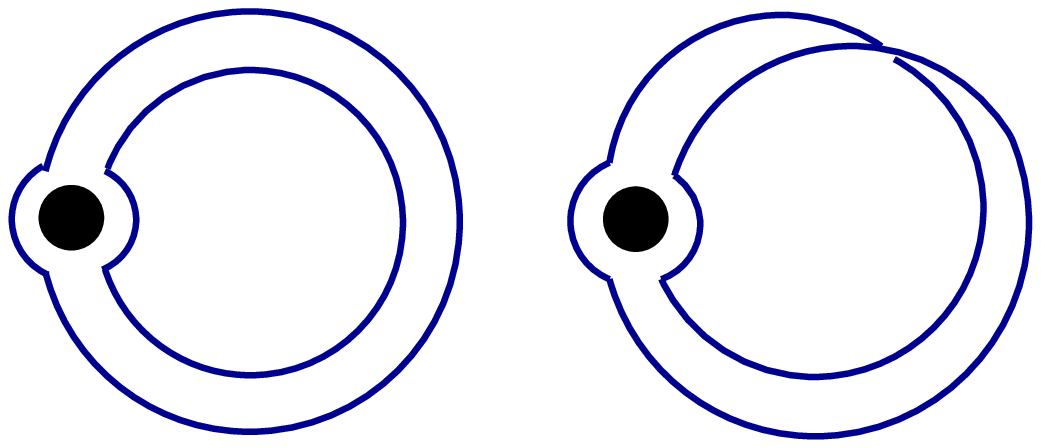}
\caption{\textsf{{\footnotesize 1-loop diagrams contributing to 
second order in $S_i$ and first order in $S_0$.}}}
\label{fig1}
\end{center}
\end{figure}

The $\cO(S^2)$ contribution to $\psiisqs$
is independent of the matter content
and comes from the first diagram in Fig.~1:
\beq
\gs \psiisqs = {S_i^2 \over \Wpp}  + \cO(S^3),
\qquad\qquad
\Wpp = \alpha  \prod_{j\neq i} (e_i-e_j) \, .
\eeq
Only theories with an antisymmetric hypermultiplet
have a $\psizsqrp$ term,
and the $\cO(S)$ contribution comes from the second diagram in Fig.~1:
\beq
\psizsqrp =  \frac{S_0}{2 \Wzpp } + \cO(S^2) \, .
\eeq
These are the only terms contributing to the one-instanton prepotential,
which becomes
\beq
\label{eq:univ}
2 \pi i \Lam^\bet \cFone
=  \sum_i  \frac{\vev{S_i}}{\Wpp} + \frac{\vev{S_0}}{\Wzpp} 
\eeq
where the second term is only present 
in theories with one antisymmetric hypermultiplet (see below
for theories with two antisymmetric hypermultiplets).
Only the $\cO(\Lam^\bet)$ contributions
to $\vev{S_i}$ and $\vev{S_0}$ 
are retained in eq.~(\ref{eq:univ}),
and we may replace $e_i$ with $a_i$ 
since the difference between these is higher order in $\Lam$.

The fact that the one-instanton prepotential takes the universal 
form (\ref{eq:univ}) was observed empirically in 
sec.~6 of ref.~\cite{Naculich:2003ka}
for a subset of the theories considered in this paper.
(This universality of form of the one-instanton
prepotential was previously recognized in 
Refs.~\cite{Ennes:1999rn,Ennes:1999fb}
based in calculations in SW theory.)

While the form (\ref{eq:univ})  is universal,
the specific expressions for $\cFone$ (as functions of $a_i$)
differ among the theories because the
extrema $\vev{S_i}$ (and $\vev{S_0}$, if present) of $\Weff$
depend on the matter content. 
We now evaluate (\ref{eq:univ}) explicitly for the four classes 
of asymptotically-free $\U(N)$ theories listed at the
beginning of sec.~2,
as well as for pure $\cN=2$ $\SO(N)$ and $\Sp(N)$ gauge theory.

\bigskip
\noindent \bfa\ 
{\bf 
$\U(N)$ with $N_f$ $\Yfund$  hypermultiplets ($0 \le N_f < 2N$) 
}

\medskip \noindent
To lowest order in $\Lam$, we have \cite{Naculich:2002hr}
\beq
\vev{S_i} =  {\alpha  \prodNf (e_i + m_I) \over \Ri} \Lam^{2N-N_f}+\cdots,
\qquad\qquad
\Ri = {\Wpp \over \alpha} = \prod_{j \neq i} (e_i - e_j)\, .
\eeq
Substituting this into eq.~(\ref{eq:univ}),
we obtain
\beq
2 \pi i \cFone =  
\sum_i  {\prodNf (a_i + m_I) \over \prod_{j \neq i} (a_i - a_j)^2} 
\eeq
where we have replaced $\Ei$ by $a_i$, 
since the difference is higher order in $\Lam$.
This agrees 
with the expression (4.34) 
derived in Ref.~\cite{D'Hoker:1997nv}
and the previous matrix-model result \cite{Naculich:2002hr}.

\bigskip
\noindent \bfb \ 
{\bf
$\U(N)$ with a $\Ysymm$  hypermultiplet 
and $N_f$ $\Yfund$ hypermultiplets 
($0 \le N_f < N-2$) 
}

\medskip \noindent
A generalization of the calculation in Ref.~\cite{Naculich:2003ka}
yields 
\beq
\vev{S_i} =  { \alpha 
(e_i - \mm)^2 \prod_{j} (e_i + e_j -2 \mm) \prodNf (e_i + m_I) 
\over \Ri} \Lam^{N - 2 - N_f} + \cdots
\eeq
giving the one-instanton prepotential 
\beq
2 \pi i \cFone =  
\sum_i { 
(a_i    - \mm)^2 \prod_{j} (a_i + a_j-2\mm) \prodNf (a_i + m_I)  
\over \prod_{j \neq i} (a_i - a_j)^2} 
\eeq
where again we have replaced $\Ei$ with $a_i$.
This agrees (up to a redefinition of the mass)
with the one-instanton prepotential (eq. 35) 
obtained for this theory in Ref.~\cite{Ennes:1998ve}.
This result was obtained previously 
using the matrix model in the $N_f=0$  case
\cite{Naculich:2003ka}.

\bigskip
\bigskip
\noindent \bfc\ 
{\bf
$\U(N)$ with an $\Yasymm$ hypermultiplet and 
$N_f$ $\Yfund$ hypermultiplets  ($0 \le N_f < N+2$)
}

\medskip \noindent
A generalization of the calculation in Ref.~\cite{Naculich:2003ka} yields
\bea
\vev{S_i} &=&  {\alpha \prod_{j} (e_i + e_j -2 \mm)\prodNf (e_i + m_I)  
\over (e_i   -   \mm)^2 \Ri} \Lam^{N+2-N_f} + \cdots, \non\\
\vev{S_0} &=&-\, { 2\Wzpp \prodNf  (\mm+m_I) \over 
              \prod_j (e_j-\mm)} \Lam^{N+2-N_f} + \cdots
\eea
which implies 
\beq
2 \pi i \cFone =  
\sum_i {\prod_{j} (a_i + a_j -2\mm) \prodNf (a_i + m_I)  
\over (a_i   -  \mm)^2 \prod_{j \neq i} (a_i-a_j)^2 } 
\ -\  {2 \prodNf (m+m_I)  \over \prod_j (a_j   -   \mm)}
\eeq
in agreement with the one-instanton prepotential (eq.~47) 
obtained for this theory in Ref.~\cite{Ennes:1998ve}.
This result was obtained previously using the matrix model 
for the $N_f=0$  case \cite{Naculich:2003ka}.

\bigskip
\noindent \bfd  \ 
{\bf
$\U(N)$ with two $\Yasymm$  hypermultiplets and 
$N_f$ $\Yfund$ hypermultiplets 
($0 \le N_f < 4$)
}

\medskip \noindent
The SW curve for this theory is known from M-theory only 
approximately \cite{Ennes:1999rn,Ennes:1999fb}
or in a form \cite{Ksir:2002}
in which explicit calculations of the prepotential
are not yet practicable,
and so we cannot state with complete confidence that 
the renormalization group equation (\ref{eq:RGfull}) 
holds in this case.
If, however, we assume that it remains valid, 
the arguments of sec.~\ref{sec:newstuff} of this paper show that
\beq
2 \pi i \Lam^\bet \cFone
=  \sum_i  \frac{\vev{S_i}}{\Wpp}
+ \frac{\vev{S_0}}{\Wzpp}  + \frac{\vev{S'_0}}{\Wnot''(\mm')}
\eeq
where, as explained in sec.~\ref{sec:MMapproach},  
the matrix model is evaluated
perturbatively about the extremum
$  \Phi_0 = \diag 
( \mm \1_{M_0}, \mm' \1_{M'_0}, e_1 \1_{M_1}, \ldots )$.
Generalizing the calculations of Ref.~\cite{Naculich:2003ka} to this
case yields:
\bea
\vev{S_i} &=&  
{\alpha
\prod_{j} (e_i + e_j -2 \mm) \prod_{j} (e_i + e_j -2 \mm') \prodNf (e_i + m_I) 
\over (e_i-\mm)^2 (e_i-\mm')^2 \Ri} \Lam^{4-N_f}+\cdots, \non\\
\vev{S_0} &=& 
  - { 2 \Wzpp \prod_j (e_j + \mm - 2\mm')   \prodNf  (\mm+m_I)
\over  (\mm - \mm')^2  \prod_j (e_j - \mm)  } \Lam^{4-N_f}+\cdots, \\
\vev{S'_0} &=&  
  -{ 2 {\Wnot''(\mm')} \prod_j (e_j + \mm' - 2\mm)   \prodNf  (\mm'+m_I)
\over  (\mm - \mm')^2  \prod_j (e_j - \mm')  } \Lam^{4-N_f}+\cdots, \non
\eea
and thus,
\bea
&& 2 \pi i \cFone =  
\sum_i 
{\prod_{j} (a_i + a_j -2\mm)  \prod_{j} (a_i + a_j -2\mm') \prodNf (a_i + m_I)  
\over 
(a_i   -  \mm)^2 (a_i   -  \mm')^2 \prod_{j \neq i} (a_i-a_j)^2 }  \\
&& \ - \  {2 \prod_j (a_j + \mm - 2\mm')   \prodNf  (\mm+m_I)
\over  (\mm - \mm')^2  \prod_j (a_j - \mm)  } 
\ - \  {2 \prod_j (a_j + \mm' - 2\mm)   \prodNf  (\mm'+m_I)
\over  (\mm - \mm')^2  \prod_j (a_j - \mm')  } \non
\eea
This is precisely what was conjectured\footnote{up to an irrelevant
sign change for $\mm$ and $\mm'$} in Ref.~\cite{Ennes:1999rn}.

\bigskip
\noindent \bfe  \ 
{\bf Pure $\cN=2$ $\SO(N)$ and $\Sp(N)$ gauge theories}

\medskip \noindent
Although the focus is on $\U(N)$ gauge theories in this paper,
the expression (\ref{eq:univ})  for the one-instanton
prepotential also applies to $\cN=2$ $\SO(N)$ and $\Sp(N)$ gauge 
theories \cite{AhnNam:2003}.
For $\SO(N)$, there is no glueball field 
$S_0$ \cite{Intriligator:2003xs},
and the leading contribution to $\vev{S_i}$ is
\beq
\vev{S_i} \propto  \left\{
\begin{array}{ll}
(\alpha e_i^4 / \Ri)  \Lam^{2N-4} & \mbox{ for even $N$}, \\
(\alpha e_i^2 / \Ri)  \Lam^{2N-4} & \mbox{ for odd $N$},
\end{array}
\right.
\qquad\qquad  \Ri = \prod_{j \neq i} (e^2_i - e^2_j)\, .
\eeq
Substituting this into eq.~(\ref{eq:univ}),
where now $\Wnot'(z) = \alpha z \prod_{i=1}^r (z^2-e^2_i)$,
with $r$ the rank of the group, 
yields a one-instanton prepotential in agreement 
with Refs.~\cite{Chan:1999gj,whitham,D'Hoker:1997mu}.

For $\Sp(N)$, on the other hand, 
the glueball field $S_0$ must be included,
and in fact gives the leading contribution:
$\vev{S_0} \propto \alpha \Lam^{N+2}$.
The other vevs are subleading:   $\vev{S_i} = \cO(\Lam^{2N+4})$.
Substituting $\vev{S_0}$ into eq.~(\ref{eq:univ}), 
with $m=0$ and 
$\Wnot'(z) = \alpha z \prod_{i=1}^r  (z^2-e^2_i)$, where $r=N/2$,
yields a one-instanton prepotential in agreement with
the Seiberg-Witten result \cite{Ennes:1999fb}.
(If massless fundamentals are present in the theory,
the $\cO(\Lam^{N+2})$ contribution to $\vev{S_0}$ vanishes
and the leading nonperturbative contribution to the prepotential
is $\cO(\Lam^{2N+4})$ \cite{Chan:1999gj,whitham,D'Hoker:1997mu}.)

\section{Matrix-model evaluation of  $\cFtwo$}

In this section, we evaluate the two-instanton contribution
to the prepotential for the pure $\cN=2$ $\U(N)$ theory.
While this computation would be quite lengthy 
(involving the three-loop free-energy)
using the methods of Ref.~\cite{Naculich:2002hi}, 
it is considerably simplified through the use of 
eqs.~(\ref{eq:threeterms})--(\ref{eq:threetermsMM}).

The topological expansion has no disk or $\rp$ contributions,
nor does $\cG_2$ contribute, 
so eqs.~(\ref{eq:threeterms})--(\ref{eq:threetermsMM}).
simplify to
\beq
\label{eq:pure}
\insttwo
= \half \sumN  
\Bigg[  \ddSj \gs \psiisqs     
- \Bigg( \ddSj \gs \tads  \Bigg)^2 \ \Bigg]  \bigvevS \, .
\eeq

\begin{figure}[htbp]
\begin{center}
\includegraphics[width=12cm,height=5.0cm]
{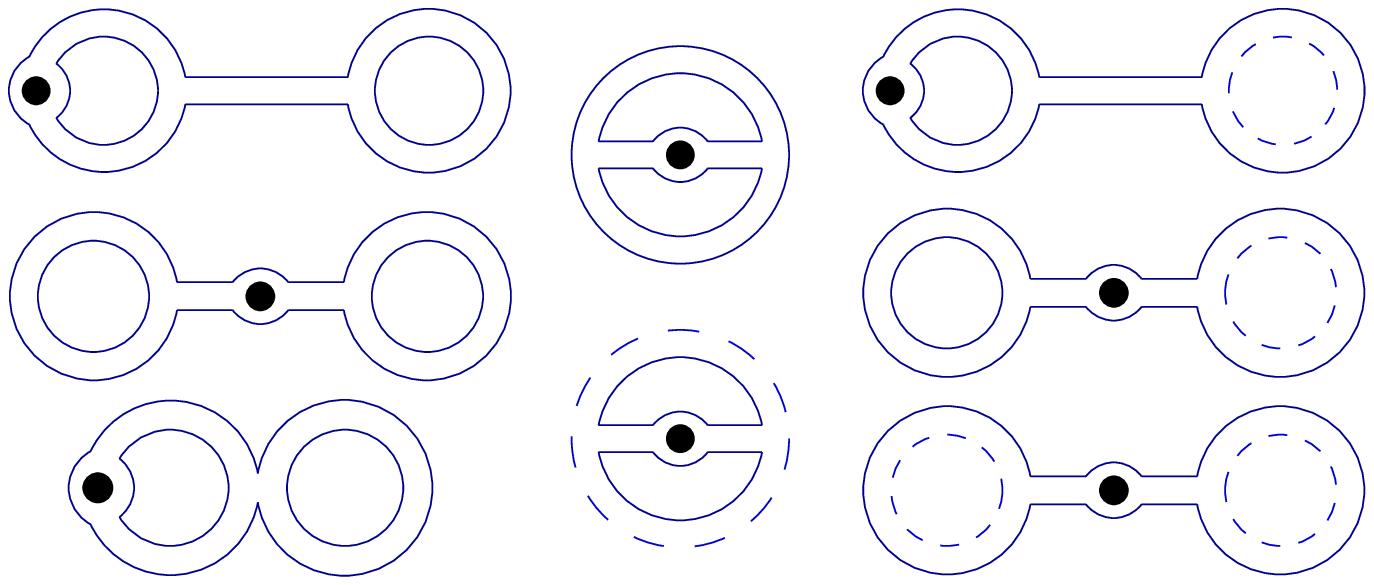}
\caption{\textsf{{\footnotesize 2-loop diagrams contributing to 
third order in $S$. 
Solid + dashed lines correspond to ghost propagators.}}}
\label{fig2}
\end{center}
\end{figure}

To evaluate the r.h.s.~to two-instanton $(\cO(\Lam^{4N}))$ accuracy, 
we only require the tadpole $\tads$
to one-loop accuracy (eq.~6.1 of Ref.~\cite{Naculich:2002hi})
\beq 
\gs \tads = {1 \over \alpha} 
\sum_{j\neq i}
\bigg[-\frac{S_i^2 }{R_i \eij}
+ 2\frac{S_i S_j}{R_i \eij} \bigg] + \cO(S^3) 
\eeq
where $\eij = e_i - e_j$.
The $\cO(S^2)$ and $\cO(S^3)$ contributions to $\psiisqs$
are obtained from the diagrams in Figs.~1 and 2, respectively
\bea
\gs \psiisqs &=& 
{S_i^2 \over \alpha \Ri} 
 +
{1 \over \alpha^2}
\Bigg[  
   \sumk {S_i^3\over \Ri^2 \eik^2 } 
+  {3  S_i^3\over \Ri^2} \bigg( \sumk {1 \over \eik} \bigg)^2
\\
&& -  \sumk {8 S_i^2 S_k \over \Ri^2 \eik}  \sumell {1\over\eil}
     +   \sumk \sumell {4 S_i S_k S_\ell \over \Ri^2 \eik \eil} 
     -   \sumk {2 S_i^2 S_k \over \Ri^2 \eik^2 } 
\Bigg] + \cO(S^4) \, .
\non
\eea
Substituting these expressions into eq.~(\ref{eq:pure}) gives
\bea
\label{eq:twoinstS}
\insttwo &=& 
\sum_i {\vev{S_i} \over \alpha \Ri} 
+
{1 \over \alpha^2}   \sum_i 
\Bigg[  
  {\vev{S_i}^2 \over 2 \Ri^2} \sumk {1 \over \eik^2} 
\\ 
&&
+ {\vev{S_i}^2  \over 2 \Ri^2} \bigg( \sumk {1 \over \eik} \bigg)^2
- \sumk \sumell { 4 \vev{S_i} \vev{S_k} \over \Ri^2 \eik \eil }
- \sumk {2 \vev{S_i} \vev{S_k} \over \Ri^2 \eik^2 } 
\Bigg] + \cO(S^3)  \, .
\non
\eea
The perturbative evaluation of $\vev{S_i}$ 
is given in eq.~(4.17) of Ref.~\cite{Naculich:2002hi}:
\bea
\label{Svev}
\vev{S_i}= 
{\alpha \Lam^{2N} \over \Ri}  
&+& {\alpha \Lam^{4N}}
 \Bigg[
  \frac{1}{2 \Ri^3} \sumk \frac{1}{\eik^2} 
+ \frac{2}{\Ri^2} \sumk \frac{1}{\Rk \eik^2}
\\
&+&
 \frac{2}{\Ri}  \sumk \frac{1}{\Rk^2\eik^2} 
+ \frac{3}{2 \Ri^3} \bigg( \sumk {1\over \eik} \bigg)^2
+ \frac{4}{\Ri}  \sum_{k \neq i}  
\sum_{\ell \neq k} \frac{1}{\Rk R_\ell \eik \ekl} 
\Bigg] + \cO(\Lam^{6N}) \, .
\non
\eea
After substituting $\vev{S_i}$ into eq.~(\ref{eq:twoinstS})
we rewrite the expression in terms of $a_i$ using
eq. (6.2) of Ref.~\cite{Naculich:2002hi}
\beq
e_i = a_i +  {2 \Lam^{2N}\over \Ri^2}\sum_{j\neq i}\frac{1}{\aij}
+ \cO(\Lam^{4N})\,,
\eeq
where $\aij = a_i - a_j$,
to find
\bea
\label{eq:twoinstL}
\insttwo&=& 
\sum_i {\Lam^{2N}  \over \Ri^2} 
+ \Lam^{4N} \sum_i \Bigg[  
  {1 \over \Ri^4} \sumk {1 \over \aik^2} 
\\
&&+ {2 \over \Ri^2} \sumk {1 \over \Rk^2 \aik^2 } 
- {2 \over \Ri^4} \bigg( \sumk {1 \over \aik} \bigg)^2 
+{4 \over \Ri^2} \sumk {1 \over \Rk^2 \aik} \sum_{\ell \neq k}  {1 \over \akl}
\non\\
&&
+{4 \over \Ri^2} \sumk {1 \over \Rk \aik} 
\sum_{\ell \neq k}  {1 \over R_\ell \akl}
-{4 \over \Ri^3} \sumk {1 \over \Rk \aik} \sum_{\ell \neq i}  {1 \over \ail}
\Bigg] + \cO(\Lam^{6N}) \non
\eea
where here and below,
$\Ri$ is redefined as $\prod_{j \neq i} (a_i - a_j)$.
Finally, using identity (4.21) of Ref.~\cite{Naculich:2002hi}
\beq
\label{id}
\sumk {1 \over \Rk \aik} = -\  {1\over \Ri} \sumk {1\over \aik}
\eeq
we obtain the one- and two-instanton contributions to the 
prepotential for pure $\cN=2$ $\U(N)$ gauge theory
\bea
\label{eq:final}
2 \pi i \cFone &=& \sum_i  {1 \over \Ri^2}\, , \non\\
2 \pi i \cFtwo &=&  \sum_i \Bigg[
  {1 \over 2 \Ri^4} \sumk {1 \over \aik^2} 
+ {1 \over \Ri^2} \sumk {1 \over \Rk^2 \aik^2 } 
+ {1\over \Ri^4} \bigg( \sumk {1 \over \aik} \bigg)^2
\Bigg] \, .
\eea
This result is precisely in agreement with 
eq.~(4.34) in Ref.~\cite{D'Hoker:1997nv}.

It would be straightforward to use eq.~(\ref{eq:threetermsMM}) to 
compute the two-instanton prepotential for the other
theories considered in this paper.

\section{Concluding remarks}

Certain results of this paper deserve to be emphasized. 
By making use of the renormalization group equations, 
a general expression for the sum of the instanton 
contributions to the prepotential was obtained 
in eqs.~(\ref{eq:threeterms})--(\ref{eq:threetermsMM})
in terms of matrix-model quantities. 
The terms in eq.~(\ref{eq:threetermsMM})
can be computed order-by-order in matrix-model perturbation theory, 
simplifying the evaluation of the prepotential within this setup. 
In particular, a direct computation in matrix-model perturbation theory 
associates a $(\kk+1)$--loop perturbative calculation 
with the $\kk$--instanton term of the prepotential, 
while the method developed in this paper requires only 
a $\kk$--loop calculation to obtain the $\kk$--instanton prepotential. 
This improvement is of most practical use for low instanton number, 
as illustrated by our one- and two-instanton results. 
The exact results (\ref{eq:threeterms})--(\ref{eq:threetermsMM})
may also lead to new insights into the matrix model. 

On the other hand, it should be pointed out that
the expression obtained in this paper
is not a pure matrix-model result,
as it relies on the RG equation satisfied by the prepotential, 
a result derived within Seiberg-Witten theory
using knowledge of the SW curve and differential.

A striking result of this paper is the universal form 
of the one--instanton correction to the prepotential (\ref{eq:univ}), 
where the dependence on the matter content appears only in 
the specific expressions for $\vev{S_i}$ and $\vev{S_0}$. 
This form applies to all aymptotically-free $\U(N)$ gauge theories, 
and to others as well (e.g., to pure $\cN=2$ $\SO(N)$ and $\Sp(N)$ theories).
In fact these regularities were previously noted in 
Refs.~\cite{Ennes:1999rn,Ennes:1999fb} 
using the Seiberg--Witten approach to the computation of the 
prepotential, 
where the quantity $S_i(x)$ 
in Refs.~\cite{Ennes:1999rn,Ennes:1999fb} 
is related to $\vev{S_i}$ of this paper 
by $S_i(e_i)={\vev{S_i}}/\Wpp$, 
with a similar expression for $\vev{S_0}$.

\section*{Acknowledgments}

We would like to thank Freddy Cachazo, Anton Ryzhov, Cumrun Vafa,
and Niclas Wyllard for useful conversations. 
HJS would like to thank the string theory group and Physics 
Department of Harvard University for their hospitality extended 
over a long period of time.   
SGN would like to acknowledge the generosity of the 2003 Simons Workshop in 
Mathematics and Physics, where some of this work was done.

\setcounter{equation}{0}
\def\theequation{A.\arabic{equation}}
\section*{Appendix}

In Ref.~\cite{Landsteiner:1998pb},
the cubic curves for $\cN=2$ $\SU(N)$ gauge theory
with one $\Yasymm$ (or one $\Ysymm$) 
and $N_f$ $\Yfund$ hypermultiplets 
were derived via M-theory.
In this appendix, we will transform these curves
into a different set of cubic curves, 
which are the ones that arise in the matrix-model approach.
We begin with the curve obtained in Ref.~\cite{Landsteiner:1998pb}
for the $\cN=2$ $\SU(N)$ gauge theory
with one $\Yasymm$ and $N_f$ $\Yfund$ hypermultiplets:
\bea
\label{eq:LLLcurve}
0 &=&  y^3 + 
\bigg[P(z) + {B \over z-m} + {3A \over (z-m)^2} \bigg] \, y^2  
\\ && 
+  {\Lam^{\bet} j(z) \over (z-m)^2}
\bigg[P(2m-z) - {B \over z-m} + {3A \over (z-m)^2} \bigg]\,  y 
+ {\Lam^{3 \bet} j(z)^2 j(2m-z)  \over (z-m)^6}
\non
\eea
where
\beq
\label{eq:ABdef}
P(z) = \prod_{i=1}^N (z-e'_i), \qquad
j(z) = \prod_{I=1}^{N_f}  (z+m_I),  \qquad
\bet = N + 2 - N_f
\eeq
with $A$ and $B$ to be specified below.
The classical moduli $e'_i$ appearing in the curve
may differ at $\cO(\Lam^\bet)$ from the moduli $e_i$
used in the matrix-model approach.
The curve (\ref{eq:LLLcurve}) lives in the space described by 
\beq
\label{eq:space}
x y = \Lam^{2 \bet} \, { j(z) j(2m-z) \over (z-m)^4}
\eeq
modded out by the involution
$x \leftrightarrow y$, $z \to 2m -z $. 

To transform this curve, 
one first defines the variable (invariant under the involution)
\beq
\label{eq:defu}
u =\,  -\,  x\,  -\,  y \,  -\,  {2 \Lam^\bet j(m) \over (z-m)^2}\,.
\eeq
Equation (\ref{eq:defu}) may be rewritten, using (\ref{eq:space}),  as
\bea
\label{eq:ysq}
y^2 &=& \, -\,  u y \, -\,  {2 \Lam^\bet j(m) \over (z-m)^2} \,\, y 
\, -\,  {\Lam^{2 \bet} j(z) j(2m-z) \over (z-m)^4} \,,
\\
\label{eq:yinv}
{1\over y} &=&  \, - \, {(z-m)^4 \over \Lam^{2 \bet} j(z) j(2m-z) } 
\left[ 
 u \, +\, y  \,+\, {2 \Lam^\bet j(m) \over (z-m)^2} 
\right] \,. 
\eea
Next divide eq. (\ref{eq:LLLcurve}) by $y$, 
and use eqs.~(\ref{eq:ysq})  and (\ref{eq:yinv}) to obtain
\bea
\label{eq:y}
y &=&\,  -\,  {\Lam^\bet j(z) \over (z-m)^2}
\left[  u \, -\,  \pi (2m-z) \over   u \, - \, \pi (z) \right] \,,
\\
\pi (z)  &\equiv &
P(z) \,+ \, {B \over z-m} 
\, +\,  {3 A -  \Lam^\bet \left[ j(z) + 2 j(m) \right]  \over (z-m)^2 } \,.
\eea
Substituting (\ref{eq:y}) into eq.~(\ref{eq:ysq}), we obtain 
the transformed cubic curve\footnote{\label{foot:invt}
This curve is invariant under $z \to 2m-z$.
The actual SW curve is the quotient of this curve by $\Z_2$.} 
\bea
\label{eq:anticurve}
&& u \left[ u- \pi(z)\right] \left[u - \pi(2m-z) \right] = 
{\Lam^\bet \over (z-m)^2} \Big\{ {j(2m-z)} \, \left[u - \pi(z) \right]^2 \non\\
&&\qquad + {j(z)} \, \left[u - \pi(2m-z) \right]^2
- {2 j(m)} \, \left[u - \pi(z) \right]\left[u - \pi(2m-z)\right] \Big\}\,.
\eea
Now, provided that the parameters $A$ and $B$ in the curve (\ref{eq:LLLcurve})
are given by
\beq
A = \Lam^\bet j(m),\qquad B = \Lam^\bet {\D j \over \D z} (m)
\eeq
eq.~(\ref{eq:anticurve}) may be written as
\beq
\label{eq:xformanti}
u  \left[u- P(z)\right] \left[u - P(2m-z)\right]   = r_1 (z) u\,  -\,  t_1 (z)
\eeq
where $r_1(z)$ and $t_1(z)$ are polynomials
(whose explicit forms are not particularly enlightening).
Indeed, the requirement that these be polynomials 
was used in Ref.~\cite{Landsteiner:1998pb}
to determine the values of $A$ and $B$.
Equation (\ref{eq:xformanti}) is the form of the curve 
that arises within the matrix-model approach to this theory
\cite{Naculich:2003ka,Klemm:2003cy,Naculich:2003cz}.

When $\Lam \to 0$, $r_1(z)$ and $t_1(z)$ vanish, and the curve
(\ref{eq:xformanti}) has singular points at the roots of
$P(z)$, $P(2m-z)$, and $P(z)-P(2m-z)$.
When $\Lam \neq 0$, the curve is deformed such that
the first two sets of singular points (at $z=e'_i$ and $z=2m-e'_i$)
open up into branch cuts (on different sheets), 
but the singularities at the roots of $P(z)-P(2m-z)$ remain,
{\it except} for the one at $z=m$, 
which also opens up into a branch cut.
Thus, on the top sheet ($u \approx P(z)$) of the Riemann surface,
there are branch cuts near $z=e'_i$ and $z=m$, 
as was assumed in sec.~\ref{sec:newstuff}.

The cubic curve for the
$\cN=2$ $\SU(N)$ gauge theory
with one $\Ysymm$ and $N_f$ $\Yfund$ hypermultiplets 
is \cite{Landsteiner:1998pb}
\beq
\label{eq:LLLsymcurve}
0 =  y^3 + P(z)  \, y^2  
+  {\Lam^{\bet} (z-m)^2 j(z) } P(2m-z) \,  y 
+ \Lam^{3\bet} (z-m)^6 j(z)^2 j(2m-z)  
\eeq
(where now $\bet = N-2-N_f$) which lives in a space described by
\beq
x y =  \Lam^{2\bet} (z-m)^4  j(z) j(2m-z) 
\eeq
modded out by the involution
$x \leftrightarrow y$, $z \to 2m -z $. 
To transform the curve into the form obtained in the matrix model,
one defines the invariant variable
\beq
u =\,  -\,  x\,  -\,  y \,  -\,  {2 \Lam^\bet j(m) \, (z-m)^2}
\eeq
and follows the same strategy as before to obtain 
\beq
y =\,  -\,  {\Lam^\bet  (z-m)^2 j(z)  }
\left[  u \, -\,  \pi (2m-z) \over   u \, - \, \pi (z) \right],
\qquad
\pi (z)  \equiv  
P(z)    \, -\,  \Lam^\bet (z-m)^2 \left[ j(z)+ 2j(m)\right]    
\eeq
from which one may derive the transformed cubic 
curve\footnote{See footnote \ref{foot:invt}.}
\bea
&& u \left[ u- \pi(z)\right] \left[ u - \pi(2m-z) \right] = 
{\Lam^\bet  (z-m)^2} \Big\{ {j(2m-z)} \, \left[u - \pi(z) \right]^2 \non\\
&&\qquad + {j(z)} \, \left[u - \pi(2m-z) \right]^2
- {2 j(m)} \, \left[u - \pi(z) \right]\left[u - \pi(2m-z)\right] \Big\}\,.
\eea
This may be rewritten as 
\beq
\label{eq:xformsym}
u 
\left[u  - P(z)    +3 \Lam^\bet j(m) (z-m)^2 \right] 
\left[u -  P(2m-z)+ 3 \Lam^\bet j(m) (z-m)^2 \right] 
= r_1 (z) u\,  -\,  t_1 (z)
\eeq
with polynomials $r_1(z)$ and $t_1(z)$.
This is the form of the curve that arises in the matrix 
model \cite{Naculich:2003ka,Klemm:2003cy,Naculich:2003cz}.

The polynomials $r_1(z)$ and $t_1(z)$ contain factors of
$(z-m)^2$, so unlike the previous case,
the singular point $z=m$ of the curve (\ref{eq:xformsym})
remains present when $\Lam \neq 0$;
i.e. it does not open up into a branch cut.
Thus, on the top sheet ($u \approx P(z)$) of the Riemann surface,
there are branch cuts only near $z=e'_i$,
as was assumed in sec.~\ref{sec:newstuff}.

\providecommand{\href}[2]{#2}\begingroup\raggedright\endgroup
\end{document}